# Quantum encryption design overcomes Shannon's theorem to achieve perfect secrecy with reusable keys

Zixuan Hu[1] and Zhenyu Li[1,2]*

1. *Key Laboratory of Precision and Intelligent Chemistry, University of Science and Technology of China, Hefei 230026, China*
2. *Hefei National Laboratory, University of Science and Technology of China, Hefei 230088, China*
*Email: zyli@ustc.edu.cn*

**Abstract:** Shannon's perfect-secrecy theorem states that a perfect encryption system that yields zero information to the adversary must be a one-time pad (OTP) with the keys randomly generated and never reused. In this work we design the first encryption method (classical or quantum) that overcomes Shannon's theorem to achieve perfect secrecy with reusable keys. Because the mechanisms used are fundamentally quantum, Shannon's theorem remains true in the classical regime. Consequently, the quantum encryption design demonstrates decisive quantum advantage by achieving a goal impossible for classical systems. Finally, the design has major practical advantages by not requiring authentication and having silent tampering detection.

## 1. Introduction

Cryptography – the interdisciplinary study of secret communication systems in the presence of potential adversaries – is important for many modern-day applications including national security, finance, and daily communications using electronic devices. In cryptography, Shannon's perfect-secrecy theorem [1] states that the only perfectly secret encryption system is the one-time-pad (OTP) that satisfies the following three conditions: 1. the keys are generated by a truly random process; 2. the size of each key (measured in bits of information) must be greater or equal to the plaintext; 3. the key must not be reused. Shannon reached these conclusions because he defined the perfectly secret encryption system as one that reveals zero information to the adversary. This is only possible by having multiple keys generated randomly such that the probability distribution of the keys conceals the statistical patterns in the ciphertexts that would allow the adversary to deduce information on the plaintext. For this concealment to be perfect, each key must be at least the same length of the plaintext and cannot be reused – this makes the OTP impractical for most communication needs due to the difficulty in the generation, sharing, and storage of large number of keys. For this reason, most commonly used encryption methods including the symmetric Advanced Encryption Standard (AES) [2] and the asymmetric Rivest-Shamir-Adleman (RSA) protocol [3] sacrifice perfect secrecy and aim for practical secrecy [1] (i.e. unrealistically enormous computational resources are needed to break the encryption) to have the keys reusable.

With the rapid development of quantum computing and information processing techniques, quantum cryptography has made significant progress in the last few decades by using unique quantum phenomena such as superposition, entanglement, and probabilistic measurement to design novel cryptographic models [4-19]. In quantum cryptography, when the ciphertext becomes a quantum state, it cannot be read exactly but can only produce probabilistic results when measured, so the adversary generally obtains less information from a quantum ciphertext as compared to a classical one that can be read exactly. However, reusing the same quantum encryption key still in general produces distinguishable measurement statistics in the ciphertext states, which can give the adversary non-zero information on the plaintext or the key. For this reason, up to now the main conclusions of Shannon's perfect-secrecy theorem are considered true even for quantum encryption systems [12-19]. However, in recent studies on the dormant entanglement phenomenon [20, 21], multi-qubit entangled states exhibited intriguing behaviors that motivated us to reconsider Shannon's perfect-secrecy theorem and design a new quantum encryption method. In one case we noticed that a reduced density state can have certain correlations with an external system that stay fundamentally inaccessible unless all the qubits are measured in some specific basis combination. This motivated the idea that if the "specific basis combination" is the key, and the "correlations" carry the information, then only the person knowing the key can recover the information, which is otherwise "fundamentally inaccessible". In another case we noticed that two entangled qubits can have no correlation when measured in arbitrary bases. This motivated the idea that if the ciphertext state is designed with all entangled qubits having no correlation to each other when measured in arbitrary bases, then its measurement statistics would give zero information to the adversary.

In this work we present the perfect-secrecy quantum encryption (PSQE) design that overcomes Shannon's theorem by achieving perfect secrecy with reusable keys. We first present a two-phase general procedure of the PSQE, followed by a simple worked-example. We then use physical intuitions and mathematical proofs to explain how the method works, achieves perfect secrecy, and has reusable keys. We then proceed to show the mechanisms of the PSQE only work with unique properties of entanglement that are fundamentally quantum. Finally we discuss the practical advantages of the PSQE such as not requiring authentication and having silent tampering detection. The PSQE is so far as we know, the first ever encryption design (classical or quantum) that overcomes Shannon's perfect-secrecy theorem. The method not only has application in cryptography, but also demonstrates decisive quantum advantage by achieving a goal impossible for classical systems.

## 2. The perfect-secrecy quantum encryption (PSQE)

Consider the scenario with Alice the sender, Bob the recipient, and Eve the adversary. Alice and Bob pre-share an ($n$-1)-bit key $\mathbf{k} = \left( k_1, ..., k_{n-1} \right)$ that is unknown to Eve. Alice, Bob, and Eve all have the ability to manipulate qubits. Alice and Bob have two public channels, one quantum and one classical, and anything sent through these channels is subject to Eve's interception and examination. Alice and Bob aim to achieve perfectly secret communication while having the key reusable. The PSQE's two-phase general procedure is presented in Box 1.

**Box 1**

**PSQE Phase 1:**

Step 1: Alice creates the $n$-qubit GHZ (Greenberger–Horne–Zeilinger) state
$$\left| GHZ^{(n)} \right\rangle = \frac{1}{\sqrt{2}} \left( \left| 00...0 \right\rangle_{12...n} + \left| 11...1 \right\rangle_{12...n} \right).$$

Step 2: Alice applies Hadamard on all qubits in $\left| GHZ^{(n)} \right\rangle$ to produce $\left| \psi^{(n)} \right\rangle$.

Step 3: For each $q_i$ from $q_1$ to $q_{n-1}$, Alice takes a new qubit $q_{iD} = \left| 0 \right\rangle$ and applies the CNOT gate $CX_{i \to iD}$ ( $i \to iD$ means $q_i$ controls $q_{iD}$ ) to produce $\left| \psi_D^{(n)} \right\rangle$.

Step 4: For each $q_i$ from $q_1$ to $q_{n-1}$, Alice reads $k_i$, the $i^{\text{th}}$ bit of the key: if $k_i = 0$ she does nothing; if $k_i = 1$ she applies Hadamard to $q_i$.

Step 5: Alice sends each $q_i$ from $q_1$ through $q_{n-1}$ to Bob, in the order indexed by $i$.

Step 6: For each received $q_i$, Bob reads $k_i$: if $k_i = 0$ he does nothing; if $k_i = 1$ he applies Hadamard to $q_i$.

Step 7a: Alice measures $q_n$'s value in the current basis.

Step 7b: Bob measures all received qubits in the current basis, calculates the sum of all values, obtains $q_n$'s value by $q_1 \oplus q_2 \oplus ... \oplus q_n = 0$.

Alice and Bob repeat Steps 1 through 7 a number of times to generate a bit string $\mathbf{b} = \left( b_1, ..., b_m \right)$ from $q_n$'s values, where $b_j$ is the $q_n$'s value measured in the $j^{\text{th}}$ repetition. $\mathbf{b}$'s length should match the plaintext as introduced in Phase 2.

**PSQE Phase 2:**

Step 8: Alice uses $\mathbf{b}$ to perform bit-wise $\oplus$ to the plaintext of the same length, generating a classical ciphertext, and then sends it to Bob.

Step 9: Bob recovers the plaintext by reading the ciphertext and $\mathbf{b}$.

Next we illustrate the procedure in Box 1 with a worked example using a 4-bit key of $\mathbf{k} = \left( 0, 1, 1, 0 \right)$ and a 5-bit plaintext of $\mathbf{p} = \left( 1, 0, 1, 0, 0 \right)$. To avoid repetition, the fully-detailed step-wise description in the same format as Box 1 is shown in the Supplementary Materials (SM) Section S1. Here we present graphical illustrations with explanations.

We start at Phase 1, whose graphical illustration is shown in Figure 1.

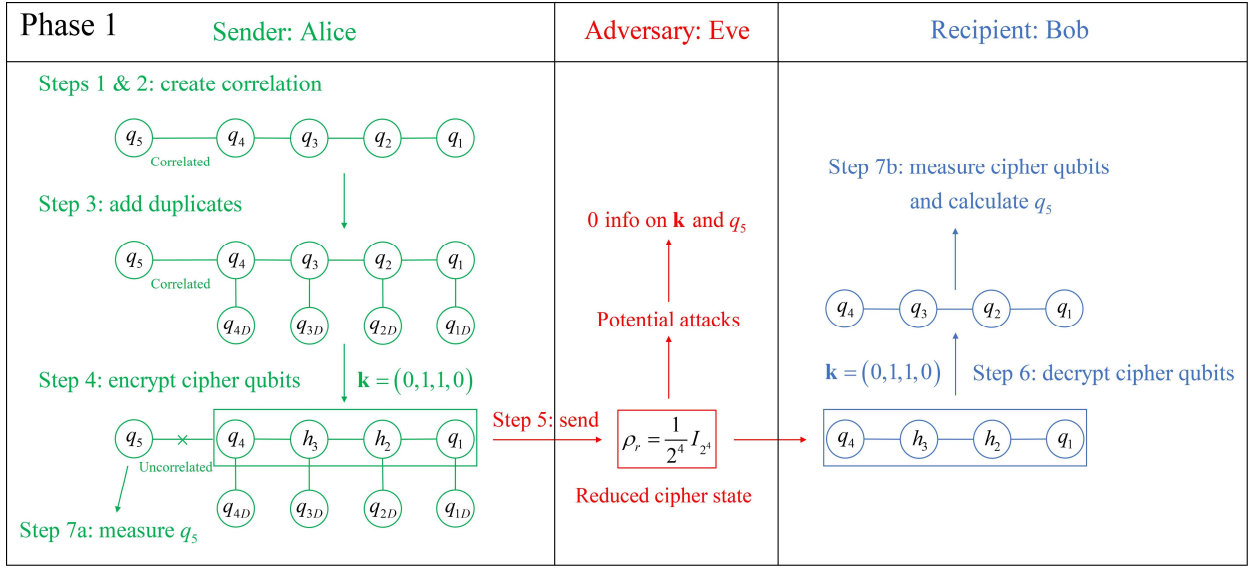

Figure 1. Graphical illustration of Phase 1 for an example using a 4-bit key $\mathbf{k} = (0,1,1,0)$. All green steps are performed locally by Alice, and all blue steps locally by Bob. Only the cipher qubits in the square box are sent to Bob, while others are kept locally with Alice. Eve can apply potential attacks in Step 5, but only on the reduced state as enclosed in the red box (see Section 3.1 below for the theorem), which yields zero information on $\mathbf{k}$. Without $\mathbf{k}$, Eve cannot reverse the encryption and obtain $q_5$'s value.

In Figure 1, Alice first creates the 5-qubit dormant entanglement state $\left|\psi^{(5)}\right\rangle$ as described in Refs. [20, 21], whose qubit values when measured in the current basis are correlated by $q_1 \oplus q_2 \oplus q_3 \oplus q_4 \oplus q_5 = 0$ (Steps 1 and 2). Alice then adds a duplicate qubit $q_{iD}$ for each cipher qubit $q_i$ from $q_1$ to $q_4$ (Step 3), and then proceeds to encrypt the cipher qubits by Hadamard-rotating them according to the key (Step 4): e.g. $\mathbf{k} = (0,1,1,0)$ means $q_2$ and $q_3$ will be Hadamard-rotated to $h_2$ and $h_3$ ("$h$" stands for "Hadamard-rotated"). These rotations cause $q_5$ to temporarily lose its correlation with the cipher qubits. The cipher qubits are sent to Bob (Step 5), who then reverses the rotations according to the key to restore $q_5$'s correlation with the cipher qubits (Step 6). Finally, Alice and Bob obtain $q_5$'s value by measuring $q_5$ and the cipher qubits respectively (Steps 7a and 7b). Eve can potentially attack in Step 5, but having access to only the cipher qubits $q_1$ through $q_4$, the reduced density state she can manipulate is $\rho_r = \frac{1}{2^4} I_{2^4}$ ( $I_{2^4}$ being the $2^4 \times 2^4$ identity matrix, see Section 3.1 for the theorem), which gives zero information on the key $\mathbf{k}$. Without the key, Eve cannot reverse the encryption and obtain $q_5$'s value by its correlation with the cipher qubits.

The quantum circuit construction of the same process as in Figure 1 is shown in Figure *2*.

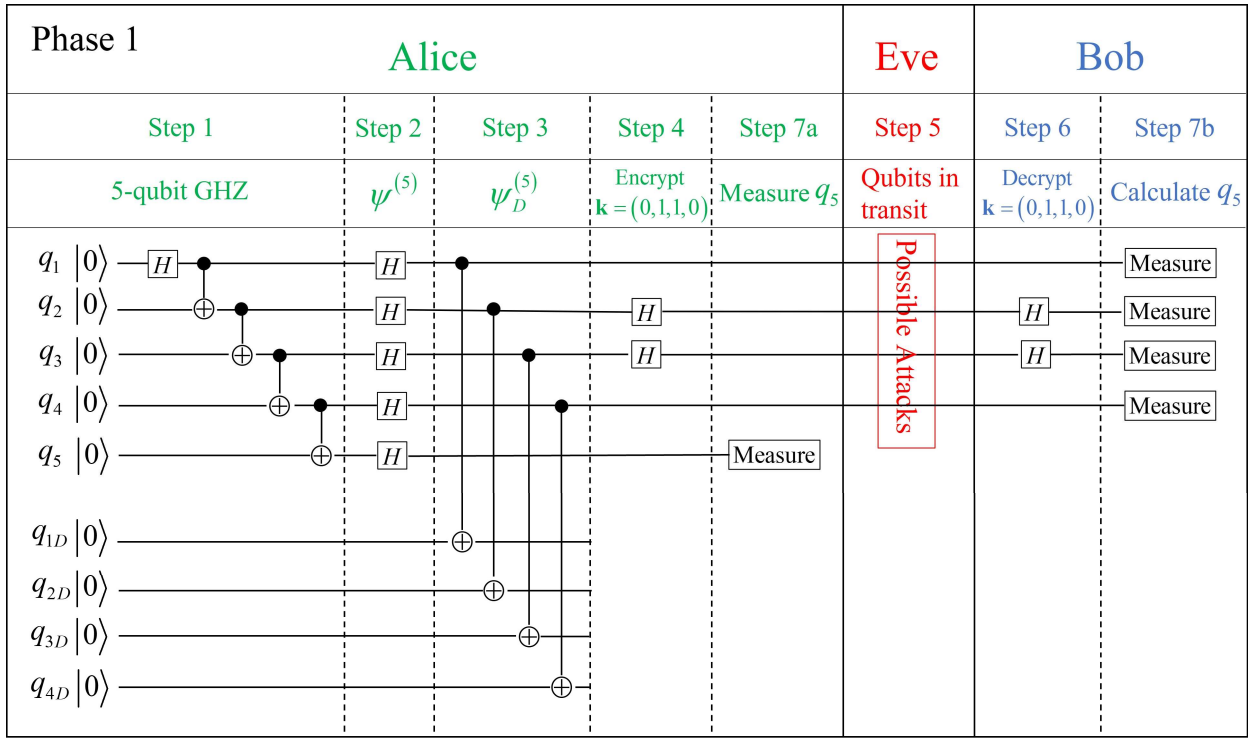

Figure 2. The quantum circuit construction of the same process as in Figure 1. All green steps are performed locally by Alice, and all blue steps locally by Bob. Eve can apply potential attacks in Step 5. Step 7a, i.e. measuring $q_5$ by Alice can happen any time after Step 2.

In Figure 2, the quantum circuit for the encryption using the 4-bit key $\mathbf{k} = (0,1,1,0)$ is constructed with 9 qubits, 8 CNOT gates, and 10 Hadamard gates (including the encryption gates by Alice and decryption gates by Bob). In general, if we want the key size to be $(n-1)$ bits, then the quantum circuit will have $(2n-1)$ qubits, $(2n-2)$ CNOT gates, and $(n+2n_k+1)$ Hadamard gates, where $n_k$ is the number of 1s in the key $\mathbf{k}$. Because the quantum circuit scales linearly with the key size, it is highly efficient when we need to have a large key.

Next we proceed to Phase 2: for this demonstration we assume Phase 1 is repeated five times to match the 5-bit length of the plaintext, generating $\mathbf{b} = (0,1,0,0,1)$.

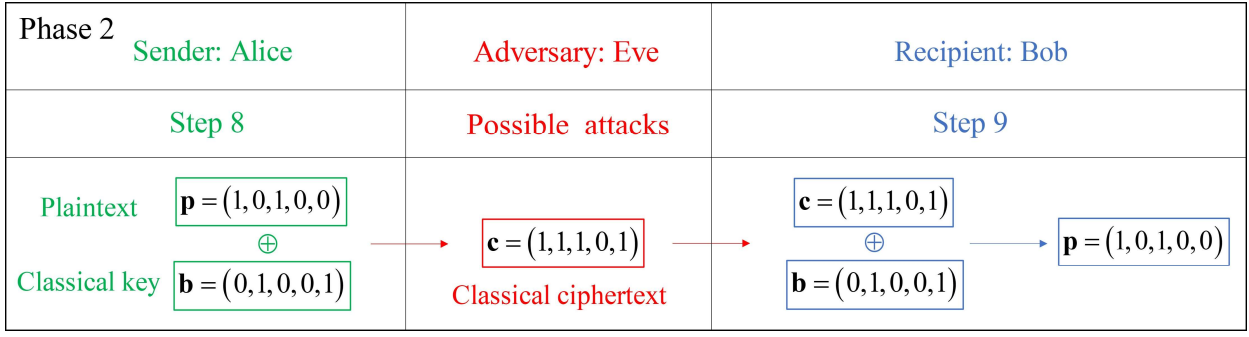

Figure 3. Graphical illustration of Phase 2 for an example using a 5-bit plaintext $\mathbf{p} = (1,0,1,0,0)$ and assuming $\mathbf{b} = (0,1,0,0,1)$. Step 8 is performed locally by Alice, Step 9 locally by Bob. Eve can attack the classical ciphertext $\mathbf{c}$ but will fail with zero information deduced.

In Figure 3, Phase 2 can be considered a classical XOR cipher using $\mathbf{b}$ as the key (assumed to be $(0,1,0,0,1)$ for the demonstration purpose). Alice performs bit-wise $\oplus$ on $\mathbf{p}$ and $\mathbf{b}$ to get the ciphertext $\mathbf{c} = (1,1,1,0,1)$, which is sent to Bob, who then performs bit-wise $\oplus$ on $\mathbf{c}$ and $\mathbf{b}$ to get the plaintext $\mathbf{p} = (1,0,1,0,0)$. Note that each $\mathbf{b}$ can be used to encrypt only one plaintext of the same length to guarantee perfect secrecy. In this phase Alice and Bob's roles are symmetric: Bob can also perform Step 8 to become the sender, with Alice performing Step 9 to become the recipient.

## 3. Analysis and discussions

### 3.1 The PSQE achieves perfect secrecy with reusable keys.

#### 3.1.1. Eve can obtain zero information on the key $\mathbf{k}$ by intercepting the cipher qubits sent to Bob in Phase 1. In Step 4, the cipher qubits $q_1$ through $q_{n-1}$ are Hadamard-rotated according to $\mathbf{k}$, so they do carry information on $\mathbf{k}$. Eve can intercept these qubits and try to extract this information by analyzing the statistical pattern in them. However, Eve can obtain zero information without accessing qubits other than $q_1$ through $q_{n-1}$ due to the following theorem:

**Theorem 1**: In $\left| \psi_D^{(n)} \right\rangle$, the reduced density state of the subsystem of $q_1$ through $q_{n-1}$ is

$$\rho_r^{(n-1)} = \rho_{1\ldots n-1} = Tr_{1D\ldots(n-1)D,n} \left( \left| \psi_D^{(n)} \right\rangle \left\langle \psi_D^{(n)} \right| \right) = \frac{1}{2^{n-1}} I_{2^{n-1}} \tag{1}$$

where we have traced away the qubits $q_{1D}$ through $q_{(n-1)D}$, and $q_n$ from $\left| \psi_D^{(n)} \right\rangle$; $I_{2^{n-1}}$ is the $2^{n-1} \times 2^{n-1}$ identity matrix.

The detailed proof of Theorem 1 is in the SM Section S2. Here we emphasize that Eq. (1) only works because the duplicate qubits $q_{1D}$ through $q_{(n-1)D}$ are added in Step 3 – without them the reduced density state would not be proportional to the identity. Now in Step 4, $q_1$ through $q_{n-1}$ are rotated according to the key **k**'s bit values, but by the theorem the reduced density state of these qubits is $\rho_r^{(n-1)} = \frac{1}{2^{n-1}} I_{2^{n-1}}$, which is invariant upon arbitrary unitary transformations. Consequently, $\rho_r^{(n-1)}$ is independent of the actual **k** used, and thus Eve can obtain zero information on the key by having access to only $q_1$ through $q_{n-1}$. An important implication of this is the same key can be reused indefinitely without compromising security, because no matter which key is used and what Eve does to the cipher qubits, she always sees the same reduced density state.

**3.1.2. By measuring the cipher qubits Bob can obtain the bit-string b as given by $q_n$'s value in each repetition of Phase 1, but Eve cannot do this.** Note that $q_n$'s value is never sent by Alice, and Bob calculates it by the correlation equation $q_1 \oplus q_2 \oplus ... \oplus q_n = 0$. However, this correlation equation only works when all $q_1$ through $q_{n-1}$ are measured in the original basis of $\left| \psi^{(n)} \right\rangle$ as created in Step 2 or equivalently $\left| \psi_D^{(n)} \right\rangle$ in Step 3. Because in Step 4 Alice rotates $q_1$ through $q_{n-1}$ according to the (n-1)-bit key **k**, Bob must reverse this in Step 6 by reading **k** before he can measure the cipher qubits and calculate $q_n$'s value. Eve can intercept the qubits sent in Step 5 and try to re-establish the correlation equation $q_1 \oplus q_2 \oplus ... \oplus q_n = 0$, but without knowing **k**, she will fail for the following reasons.

Suppose we are Eve for the moment and consider the simple strategy of trying to guess **k** exactly and perform Step 6. Then even if we have guessed 1 bit wrong out of (n-1) bits, we would measure one cipher qubit in the Hadamard-rotated basis, and $q_n$'s value becomes completely random. Now suppose we no longer try to guess **k** exactly, can we find an (n-1)-qubit unitary transformation to generate a specific state on the cipher qubits, which would allow us to infer $q_n$'s value correctly with a probability greater than $\frac{1}{2}$ regardless of **k**? To answer this question we present the following theorem:

**Theorem 2**: If Eve is allowed to perform arbitrary unitary transformation before measuring the cipher qubits, then her "success probability" $P_s$ of inferring $q_n$'s value correctly when averaged over all possible key configurations satisfy:

$$\frac{1}{2} - \left( \frac{\sqrt{2}}{2} \right)^{n+1} = P_{\min} \leq P_s \leq P_{\max} = \frac{1}{2} + \left( \frac{\sqrt{2}}{2} \right)^{n+1} \qquad (2)$$

Proof: With the fully detailed proof shown in the SM Section S3, here we only briefly summarize the important steps. To start, we use physical intuitions and Bayes' rule to write $P_s$ as $P_s = P\left( \bigoplus_{i=1}^{n-1} q_i = 0 \middle| q_n = 0 \right)$, i.e. the conditional probability of $\bigoplus_{i=1}^{n-1} q_i = 0$ given $q_n = 0$ has

been measured. Then we construct Eve's effective density state $\rho_e^{(n-1)}$ of the $(n\text{-}1)$ cipher qubits, averaged over all possible key configurations. Next we show that $P_s$ is equal to $\sum_{q_1 \oplus \ldots \oplus q_{n-1} = 0} \langle q_1 \ldots q_{n-1} | U \rho_e^{(n-1)} U^\dagger | q_1 \ldots q_{n-1} \rangle$, i.e. the sum of expectation values of the observable $\rho_e^{(n-1)}$ when evaluated on some basis states rotated by an arbitrary unitary transformation $U$. Then by an extension of the variational principle, we have $\sum_{i=1}^{2^{n-2}} \lambda_i \leq P_s \leq \sum_{i=2^{n-2}+1}^{2^{n-1}} \lambda_i$, where $\lambda_1 \leq \lambda_2 \leq \ldots \leq \lambda_{2^{n-1}}$ are the eigenvalues of $\rho_e^{(n-1)}$. Finally, we calculate the eigenvalues of $\rho_e^{(n-1)}$ to reach Eq. (2).

Theorem 2 means that, no matter what unitary transformation Eve applies to the intercepted cipher qubits, her probability of inferring $q_n$'s value correctly when averaged over all possible key configurations will converge from both sides to $\frac{1}{2}$, and the convergence rate is exponential in the number of cipher qubits. From a purist's perspective, this means Eve gains non-zero information on $q_n$'s value in Phase 1 of the PSQE, which makes it not strictly having perfect secrecy. However, this information gain for Eve decreases exponentially with $n$, so it quickly becomes irrelevant for all practical matters even for a relatively small $n$. Recognizing the process of generating $q_n$ is a Bernoulli trial, the information entropy of a single trial is $H_2(P) = -P \log_2(P) - (1-P) \log_2(1-P)$. The Taylor expansion of $H_2(P)$ around $\frac{1}{2}$ has no 1$^{st}$ and 3$^{rd}$ order terms, and the 2$^{nd}$ order coefficient is $-\frac{2}{\ln(2)}$, so the entropy gain for Eve is approximately $g \leq \frac{1}{\ln(2)} \left( \frac{1}{2} \right)^n$, which decays exponentially to zero. For any application, there is always a minimum level of environmental and instrumental noise that defines a threshold, below which any potential information gain is effectively overwhelmed by the noise. Regardless of future technological advancements, this threshold cannot be reduced to zero, as we can never achieve absolute zero temperature or perfectly pure materials. Given the exponential decay of $g$ with $n$, we can always ensure that Eve's information gain falls below any desired threshold with a reasonable $n$, thereby guaranteeing that she gains effectively no information – perfect secrecy is achieved. Note that Theorem 2 means the PSQE's security in Phase 1 is proved, and not based on the assumption of the hardness of some mathematical problems like most current cryptographic models. Indeed, the result in Eq. (2) is physically supported and there is no mathematical problem to solve, so Eve's success probability cannot be improved by her having more or even infinite computational resources.

### 3.1.3. Eve can obtain zero information from the classical ciphertext sent to Bob in Phase 2.
In Step 8, Alice essentially uses the bit-string **b** generated in Phase 1 as a classical key to encrypt the plaintext. If we consider Phase 2 alone as a classical XOR cipher, then it satisfies the three previously mentioned conditions of Shannon's perfect-secrecy theorem: 1. The bit-string **b** is generated by a truly random quantum process and has the maximum entropy. 2. By design the size of **b** is equal to the plaintext. 3. **b** is never reused: for each plaintext a new **b** is generated by the truly random quantum process. Consequently, Phase 2 alone is a classical perfect-secrecy system, and Eve can obtain zero information from the classical ciphertext sent in this phase.

**3.1.4. The PSQE achieves perfect secrecy with reusable keys.** By the above discussions, Eve can obtain zero (exponentially small) information on both the key and the plaintext from the two batches of ciphertexts sent in Phase 1 and Phase 2 respectively, which cannot be improved by having more or even infinite computational resources, therefore the PSQE design achieves perfect secrecy. More importantly, in principle the same key **k** can be reused indefinitely without compromising perfect secrecy. In actual application, it may still be good practice to change the key from time to time in case it gets leaked by factors outside the encryption design (e.g. Bob defects). This is so far as we know, the first ever encryption design (classical or quantum) that overcomes Shannon's theorem by achieving perfect secrecy with reusable keys.

**3.2 Perfect secrecy with reusable keys can only be achieved by a quantum system.**

The core idea of the PSQE is in Phase 1 where exotic properties of the dormant entanglement phenomenon [20, 21] are exploited to produce the same reduced density state $\rho_r = \frac{1}{2^{n-1}} I_{2^{n-1}}$ for the subsystem of $q_1$ through $q_{n-1}$, independent from the actual **k** and $q_n$'s value (the message of this phase). Classically, suppose the ciphertext is always in a completely random mixture resembling $\rho_r = \frac{1}{2^{n-1}} I_{2^{n-1}}$, regardless of the key and the message, then Bob would also extract zero information, making the encryption method useless. Indeed, this trivial case was not considered in Shannon's original discussion of the perfect-secrecy theorem [1]. In a classical cryptosystem, the key operates on the message to produce the ciphertext, so the message, the key, and the ciphertext are all dependent on each other. However, to achieve perfect secrecy the ciphertext must not yield any information on the message to Eve, so the ciphertext-message dependence must be concealed by the randomness from the key – this is why the key can never be reused in Shannon's theorem. In contrast, in the PSQE Phase 1, $q_n$'s value (the message of this phase), the key **k**, and the cipher qubit state are all independent from each other. Firstly, the key **k** does not operate on the particular value of $q_n$ to produce a particular state for the cipher qubits $q_1$ through $q_{n-1}$. Instead, **k** is only used to specify the basis rotations to first disrupt and later restore the correlation between $q_n$ and the cipher qubits – this uses the quantum property that some correlation between qubits only works when these qubits are measured in a particular basis combination. Secondly, the cipher qubit state is always described by the same reduced density state $\rho_r = \frac{1}{2^{n-1}} I_{2^{n-1}}$, so it is independent from both $q_n$ and **k**. Consequently, Eve can obtain zero information on both $q_n$ and **k** by examining the cipher qubits, even when **k** is reused.

Note that in Step 4, the cipher qubits are indeed rotated according to $\mathbf{k}$, so it may be puzzling why Eve can extract zero information on $\mathbf{k}$ from these qubits. To understand this, we note the unique quantum behavior that certain information inside a subsystem of an entangled state may be fundamentally inaccessible to an observer restricted to the subsystem. In the examples shown in the previous dormant entanglement studies [20, 21], the knowledge of whether the subsystem can be separated into entangled pure states or not is inaccessible to an observer restricted to the subsystem. In the current PSQE design, how the cipher qubits are rotated according to $\mathbf{k}$ is knowledge inside the subsystem of these qubits, but it is fundamentally inaccessible to the restricted observer Eve who can only examine the cipher qubits. In summary, the two properties of $\mathbf{k}$ being unrelated to $q_n$'s value and $\mathbf{k}$'s information being inaccessible to Eve enable the PSQE to safely reuse $\mathbf{k}$ while maintaining perfect secrecy. The first property relies on quantum correlation being only measurable in some particular basis combination; the second property relies on quantum entanglement's unique behaviors. Consequently, perfect secrecy with reusable keys can only be achieved by a quantum system, and Shannon's theorem remains true in the classical regime. In this sense, the PSQE design has decisive quantum advantage over classical encryption methods by achieving a goal impossible in the classical regime.

### 3.3 Practical advantages of the PSQE.

In terms of classical cryptography, the PSQE Phase 1 not only works as a true random number generator (TRNG) but also shares the random number with perfect secrecy and has reusable keys. Most classical cryptographic systems use cryptographically secure pseudorandom number generators (CSPRNG) because it is easier to generate and share a CSPRNG than a TRNG [22]. However, because a CSPRNG is not truly random, it can be subject to various cryptanalytic attacks including direct cryptanalytic attacks, input-based attacks, and state compromise extension attacks [23]. The generation of $q_n$'s values in the PSQE Phase 1 is guaranteed by quantum physics to be truly random, therefore it is a TRNG automatically immune to all those attacks for a CSPRNG. More importantly, as explained in Section 3.1, by exploiting unique properties of entanglement, the PSQE Phase 1 also shares the random numbers to Bob with perfect secrecy and reusable keys. This combination of features – true randomness, perfect secrecy, and reusable keys—is unattainable in classical cryptographic systems. By having all three features, the PSQE offers the benefits of an OTP without its inherent drawbacks.

The PSQE also possesses major advantages compared to existing quantum cryptographic methods. Firstly, because Bob has a critical advantage over Eve by knowing the key $\mathbf{k}$, the PSQE does not require authentication. The PSQE's process of generating the bit-string $\mathbf{b}$ in Phase 1 can be considered as a "key generation and distribution" process that achieves a goal similar to the quantum key distribution (QKD) [4-11]. However, a unique feature of the PSQE is: without knowing $\mathbf{k}$, every bit of $\mathbf{b}$ is perfectly secure from Eve. Therefore even if Eve has successfully impersonated Bob, she can still gain zero information from Alice. In fact, having perfect secrecy, the PSQE can be used to authenticate Bob, by Bob instead of Alice performing Step 8 to send a pre-determined code word, and Alice performing Step 9 to read it. To successfully impersonate Bob, Eve would need to know both the code word and the key, which is impossible because they are pre-shared and never sent. Because the PSQE does not require authentication and can perform authentication, it is safe against man-in-the-middle (MITM) attacks [22].

Secondly, because every bit of **b** is perfectly secure from Eve, the PSQE does not rely on tampering detection for its security and does not have the "announce-compare-discard" step usually used in other quantum cryptographic methods. In the PSQE, Eve never gains any knowledge of the bases and values of Alice and Bob's measurements, and can never confirm whether she has guessed anything correctly, so there is no partial knowledge gained for her. In addition, there is also no feedback that may inform Eve on the habits of Alice and Bob and the characteristics of their equipment, when their ways of generating random numbers and manipulating qubits are not perfectly random.

Thirdly, although the PSQE does not rely on tampering detection, it indeed can detect tampering. By intercepting, manipulating and measuring the cipher qubits, Eve will have altered them and destroyed their correlation with $q_n$. So when she later sends the cipher qubits to Bob, he can no longer restore the correlation, and will have no better probability than Eq. (2) to obtain $q_n$'s value. This will become obvious when **b** grows longer so Bob can detect Eve's tampering activities. Different from the "announce-compare-discard" step usually used in other quantum cryptographic methods, tampering detection in the PSQE happens naturally and silently when Bob tries to recover the plaintext in Phase 2, without any extra procedures. In other quantum cryptographic methods, Eve can observe Alice and Bob's interaction in the "announce-compare-discard" step to confirm whether her tampering activity has been detected or not; while in the PSQE, Eve never gets any feedback and thus must assume her tampering can be detected at any moment. Despite being subtle, this kind of asymmetric knowledge in cryptography gives the legitimate parties more advantages against the adversary.

## 4. Conclusion

In this study we have designed a quantum encryption method PSQE that overcomes Shannon's theorem to achieve perfect secrecy with reusable keys. Fundamentally quantum properties of entanglement are exploited to achieve this goal that is still impossible in the classical regime. In the quantum regime, certain information of the reduced density state can hide deeply in its correlation with an external system, which can only be recovered if all qubits in the subsystem are measured in the correct bases – knowing the correct bases allows the legitimate recipient to obtain information that is fundamentally inaccessible to the adversary. This is so far as we know, the first ever encryption design (classical or quantum) that overcomes Shannon's perfect-secrecy theorem. In addition to its potential application in quantum cryptography, the PSQE also demonstrates decisive quantum advantage by achieving a goal impossible in the classical regime.

### 5. Acknowledgements

This work is supported by the Innovation Program for Quantum Science and Technology (2021ZD0303306), the Strategic Priority Research Program of the Chinese Academy of Sciences (XDB0450101), and the National Natural Science Foundation of China (22393913).

**Supplementary Materials can be found after the References.**

### References:

1. Shannon, C.E., *Communication theory of secrecy systems.* The Bell System Technical Journal, 1949. **28**(4): p. 656-715.
2. Nechvatal, J., et al., *Report on the Development of the Advanced Encryption Standard (AES).* Journal of research of the National Institute of Standards and Technology, 2001. **106**(3): p. 511-577.
3. Rivest, R.L., A. Shamir, and L. Adleman, *A method for obtaining digital signatures and public-key cryptosystems.* Commun. ACM, 1978. **21**(2): p. 120–126.
4. Gisin, N., et al., *Quantum cryptography.* Reviews of Modern Physics, 2002. **74**(1): p. 145-195.
5. Pirandola, S., et al., *Advances in quantum cryptography.* Advances in Optics and Photonics, 2020. **12**(4): p. 1012-1236.
6. Bennett, C.H. and G. Brassard, *Quantum cryptography: Public key distribution and coin tossing.* Theoretical Computer Science, 2014. **560**: p. 7-11.
7. Ekert, A.K., *Quantum cryptography based on Bell's theorem.* Physical Review Letters, 1991. **67**(6): p. 661-663.
8. Bennett, C.H., G. Brassard, and N.D. Mermin, *Quantum cryptography without Bell's theorem.* Physical Review Letters, 1992. **68**(5): p. 557-559.
9. Jennewein, T., et al., *Quantum Cryptography with Entangled Photons.* Physical Review Letters, 2000. **84**(20): p. 4729-4732.
10. Xu, F., et al., *Secure quantum key distribution with realistic devices.* Reviews of Modern Physics, 2020. **92**(2): p. 025002.
11. Yin, J., et al., *Entanglement-based secure quantum cryptography over 1,120 kilometres.* Nature, 2020. **582**(7813): p. 501-505.
12. Hu, Z. and S. Kais, *A quantum encryption design featuring confusion, diffusion, and mode of operation.* Scientific Reports, 2021. **11**(1): p. 23774.
13. Zhou, N., et al., *Novel qubit block encryption algorithm with hybrid keys.* Physica A: Statistical Mechanics and its Applications, 2007. **375**(2): p. 693-698.
14. Li, J., Z. Hu, and S. Kais, *Practical quantum encryption protocol with varying encryption configurations.* Physical Review Research, 2021. **3**(2): p. 023251.
15. Boykin, P.O. and V. Roychowdhury, *Optimal encryption of quantum bits.* Physical Review A, 2003. **67**(4): p. 042317.
16. Ambainis, A., et al. *Private quantum channels.* in *Proceedings 41st Annual Symposium on Foundations of Computer Science.* 2000.

17.  Ambainis, A. and A. Smith. *Small Pseudo-random Families of Matrices: Derandomizing Approximate Quantum Encryption*. in *Approximation, Randomization, and Combinatorial Optimization. Algorithms and Techniques*. 2004. Berlin, Heidelberg: Springer Berlin Heidelberg.

18.  Kuang, R. and N. Bettenburg. *Shannon Perfect Secrecy in a Discrete Hilbert Space*. in *2020 IEEE International Conference on Quantum Computing and Engineering (QCE)*. 2020.

19.  Lai, C.-Y. and K.-M. Chung, *Quantum encryption and generalized Shannon impossibility.* Designs, Codes and Cryptography, 2019. **87**(9): p. 1961-1972.

20.  Hu, Z. and S. Kais, *Dormant entanglement that can be activated or destroyed by the basis choice of measurements on an external system.* arXiv:2306.05517, 2023.

21.  Hu, Z. and S. Kais, *The qubit information logic theory for understanding multi-qubit entanglement and designing exotic entangled states.* arXiv:2402.15699, 2024.

22.  Katz, J. and Y. Lindell, *Introduction to Modern Cryptography, Second Edition*. 2014: Chapman & Hall/CRC.

23.  Kelsey, J., et al. *Cryptanalytic Attacks on Pseudorandom Number Generators*. in *Fast Software Encryption*. 1998. Berlin, Heidelberg: Springer Berlin Heidelberg.

## Supplementary materials for "Quantum encryption design overcomes Shannon's theorem to achieve perfect secrecy with reusable keys"

Zixuan Hu[1] and Zhenyu Li[1,2]*

1.  *Key Laboratory of Precision and Intelligent Chemistry, University of Science and Technology of China, Hefei 230026, China*
2.  *Hefei National Laboratory, University of Science and Technology of China, Hefei 230088, China*
*\*Email: zyli@ustc.edu.cn*

**Section S1. Step-wise description of the worked example as illustrated in Section 2 of the main text**:

**Start of the communication.**

**PSQE Phase 1:**

Step 1: Alice creates the 5-qubit GHZ state $\left|GHZ^{(5)}\right\rangle = \frac{1}{\sqrt{2}}\left(\left|00000\right\rangle + \left|11111\right\rangle\right)$.

Step 2: Alice applies Hadamard on all qubits in $\left|GHZ^{(5)}\right\rangle$ to produce $\left|\psi^{(5)}\right\rangle$:

$$\left|\psi^{(5)}\right\rangle = \frac{1}{4} \left\{ \begin{array}{l} \left( \begin{array}{l} \Big[ \big(|00\rangle_{12} + |11\rangle_{12}\big)|0\rangle_3 + \big(|01\rangle_{12} + |10\rangle_{12}\big)|1\rangle_3 \Big]|0\rangle_4 \\ + \Big[ \big(|00\rangle_{12} + |11\rangle_{12}\big)|1\rangle_3 + \big(|01\rangle_{12} + |10\rangle_{12}\big)|0\rangle_3 \Big]|1\rangle_4 \end{array} \right)|0\rangle_5 \\ + \left( \begin{array}{l} \Big[ \big(|00\rangle_{12} + |11\rangle_{12}\big)|0\rangle_3 + \big(|01\rangle_{12} + |10\rangle_{12}\big)|1\rangle_3 \Big]|1\rangle_4 \\ + \Big[ \big(|00\rangle_{12} + |11\rangle_{12}\big)|1\rangle_3 + \big(|01\rangle_{12} + |10\rangle_{12}\big)|0\rangle_3 \Big]|0\rangle_4 \end{array} \right)|1\rangle_5 \end{array} \right\} \quad (1)$$

Remark: In $\left|\psi^{(5)}\right\rangle$ the qubit values when measured in the current basis satisfy: $q_1 \oplus q_2 \oplus q_3 \oplus q_4 \oplus q_5 = 0$.

Step 3: For each $q_i$ from $q_1$ to $q_4$, Alice takes a new qubit $q_{iD} = |0\rangle$ and applies the CNOT gate $CX_{i \to iD}$ to produce $\left|\psi_D^{(5)}\right\rangle$:

$$\left|\psi_D^{(5)}\right\rangle = \frac{1}{4} \left\{ \begin{array}{l} \left( \begin{array}{l} \Big[ \big(|00\rangle_{12}|00\rangle_{1D2D} + |11\rangle_{12}|11\rangle_{1D2D}\big)|0\rangle_3|0\rangle_{3D} \\ + \big(|01\rangle_{12}|01\rangle_{1D2D} + |10\rangle_{12}|10\rangle_{1D2D}\big)|1\rangle_3|1\rangle_{3D} \Big]|0\rangle_4|0\rangle_{4D} \\ + \Big[ \big(|00\rangle_{12}|00\rangle_{1D2D} + |11\rangle_{12}|11\rangle_{1D2D}\big)|1\rangle_3|1\rangle_{3D} \\ + \big(|01\rangle_{12}|01\rangle_{1D2D} + |10\rangle_{12}|10\rangle_{1D2D}\big)|0\rangle_3|0\rangle_{3D} \Big]|1\rangle_4|1\rangle_{4D} \end{array} \right)|0\rangle_5 \\ + \left( \begin{array}{l} \Big[ \big(|00\rangle_{12}|00\rangle_{1D2D} + |11\rangle_{12}|11\rangle_{1D2D}\big)|0\rangle_3|0\rangle_{3D} \\ + \big(|01\rangle_{12}|01\rangle_{1D2D} + |10\rangle_{12}|10\rangle_{1D2D}\big)|1\rangle_3|1\rangle_{3D} \Big]|1\rangle_4|1\rangle_{4D} \\ + \Big[ \big(|00\rangle_{12}|00\rangle_{1D2D} + |11\rangle_{12}|11\rangle_{1D2D}\big)|1\rangle_3|1\rangle_{3D} \\ + \big(|01\rangle_{12}|01\rangle_{1D2D} + |10\rangle_{12}|10\rangle_{1D2D}\big)|0\rangle_3|0\rangle_{3D} \Big]|0\rangle_4|0\rangle_{4D} \end{array} \right)|1\rangle_5 \end{array} \right\} \quad (2)$$

Step 4: Alice reads the key $\mathbf{k} = (0,1,1,0)$ and applies Hadamard to $q_2$ and $q_3$:

$$H_2H_3\left|\psi_D^{(5)}\right\rangle = \frac{1}{4}\left\{\begin{array}{l}\left(\begin{array}{l}\left[\begin{array}{l}\left(|0\rangle_1|+\rangle_2|00\rangle_{1D2D}+|1\rangle_1|-\rangle_2|11\rangle_{1D2D}\right)|+\rangle_3|0\rangle_{3D}\\+\left(|0\rangle_1|-\rangle_2|01\rangle_{1D2D}+|1\rangle_1|+\rangle_2|10\rangle_{1D2D}\right)|-\rangle_3|1\rangle_{3D}\end{array}\right]|0\rangle_4|0\rangle_{4D}\\+\left[\begin{array}{l}\left(|0\rangle_1|+\rangle_2|00\rangle_{1D2D}+|1\rangle_1|-\rangle_2|11\rangle_{1D2D}\right)|-\rangle_3|1\rangle_{3D}\\+\left(|0\rangle_1|-\rangle_2|01\rangle_{1D2D}+|1\rangle_1|+\rangle_2|10\rangle_{1D2D}\right)|+\rangle_3|0\rangle_{3D}\end{array}\right]|1\rangle_4|1\rangle_{4D}\end{array}\right)|0\rangle_5\\+\left(\begin{array}{l}\left[\begin{array}{l}\left(|0\rangle_1|+\rangle_2|00\rangle_{1D2D}+|1\rangle_1|-\rangle_2|11\rangle_{1D2D}\right)|+\rangle_3|0\rangle_{3D}\\+\left(|0\rangle_1|-\rangle_2|01\rangle_{1D2D}+|1\rangle_1|+\rangle_2|10\rangle_{1D2D}\right)|-\rangle_3|1\rangle_{3D}\end{array}\right]|1\rangle_4|1\rangle_{4D}\\+\left[\begin{array}{l}\left(|0\rangle_1|+\rangle_2|00\rangle_{1D2D}+|1\rangle_1|-\rangle_2|11\rangle_{1D2D}\right)|-\rangle_3|1\rangle_{3D}\\+\left(|0\rangle_1|-\rangle_2|01\rangle_{1D2D}+|1\rangle_1|+\rangle_2|10\rangle_{1D2D}\right)|+\rangle_3|0\rangle_{3D}\end{array}\right]|0\rangle_4|0\rangle_{4D}\end{array}\right)|1\rangle_5\end{array}\right\} \quad (3)$$

Step 5: Alice sends each $q_i$ from $q_1$ through $q_4$ to Bob, in the order indexed by $i$.

Step 6: Bob reads the key $\mathbf{k} = (0,1,1,0)$ and applies Hadamard to $q_2$ and $q_3$ to restore $\left|\psi^{(5)}\right\rangle$.

Step 7a: Alice measures $q_5$'s value in the current basis.

Step 7b: Bob measures $q_1$ through $q_4$ in the current basis, calculates $q_5$'s value by $q_1 \oplus q_2 \oplus q_3 \oplus q_4 \oplus q_5 = 0$.

Alice and Bob repeat Steps 1 through 7 five times, to obtain a bit string $\mathbf{b}$ from $q_5$'s values, where $b_j$ is the $q_1$'s value measured in the $j^{\text{th}}$ repetition. For the demonstration purpose, suppose $\mathbf{b} = (0,1,0,0,1)$.

**PSQE Phase 2:**

Step 8: Alice uses $\mathbf{b}$ to perform bit-wise $\oplus$ to $\mathbf{p} = (1,0,1,0,0)$, generating a classical ciphertext $\mathbf{c} = (1,1,1,0,1)$, and then sends it to Bob.

   Remark: $\mathbf{b}$ can be used to encrypt only one plaintext of the same length.

Step 9: Bob recovers the plaintext $\mathbf{p} = (1,0,1,0,0)$ by reading Alice's information and $\mathbf{b}$.

**End of the communication.**

**Section S2. Proof for Theorem 1 in Section 3.1 of the main text.**

**Theorem 1**: In $\left|\psi_D^{(n)}\right\rangle$, the reduced density state of the subsystem of $q_1$ through $q_{n-1}$ is

$\rho_r^{(n-1)} = \rho_{1\ldots n-1} = Tr_{1D\ldots(n-1)D,n}\left(\left|\psi_D^{(n)}\right\rangle\left\langle\psi_D^{(n)}\right|\right) = \frac{1}{2^{n-1}}I_{2^{n-1}}$, where $I_{2^{n-1}}$ is the $2^{n-1} \times 2^{n-1}$ identity

matrix.

To prove the theorem we first consider the following properties of the dormant entanglement state.

**Property 1**: The $(n+1)$-qubit dormant entanglement state $\left|\psi^{(n+1)}\right\rangle$ is related to the

$n$-qubit dormant entanglement state $\left|\psi^{(n)}\right\rangle$ by: $\left|\psi^{(n+1)}\right\rangle = \frac{1}{\sqrt{2}}\left(\left|\psi^{(n)}\right\rangle\left|0\right\rangle_{n+1} + \left|\tilde{\psi}^{(n)}\right\rangle\left|1\right\rangle_{n+1}\right)$,

where $\left|\tilde{\psi}^{(n)}\right\rangle = X_n\left|\psi^{(n)}\right\rangle$ with $X_n$ negating the value of $q_n$.

Proof: This can be proved by studying the structure of $\left|\psi^{(n)}\right\rangle$, which is created in Step 2 by applying Hadamard on all qubits of the $n$-qubit GHZ state.

By definition:

$$\left|\psi^{(n)}\right\rangle = H_1\ldots H_n\left|GHZ^{(n)}\right\rangle \quad\text{and}\quad \left|\psi^{(n+1)}\right\rangle = H_1\ldots H_n H_{n+1}\left|GHZ^{(n+1)}\right\rangle \qquad (4)$$

$\left|GHZ^{(n+1)}\right\rangle$ is created from $\left|GHZ^{(n)}\right\rangle$ by:

$$\left|GHZ^{(n+1)}\right\rangle = CX_{n\to n+1}\left|GHZ^{(n)}\right\rangle \otimes \left|0\right\rangle_{n+1} \qquad (5)$$

So we have:

$$\begin{aligned}
\left|\psi^{(n+1)}\right\rangle &= H_1\ldots H_n H_{n+1}\left|GHZ^{(n+1)}\right\rangle \\
&= H_1\ldots H_n H_{n+1} CX_{n\to n+1}\left|GHZ^{(n)}\right\rangle \otimes \left|0\right\rangle_{n+1} \\
&= H_1\ldots\left(H_n H_{n+1} CX_{n\to n+1} H_{n+1} H_n\right) H_n H_{n+1}\left|GHZ^{(n)}\right\rangle \otimes \left|0\right\rangle_{n+1} \\
&= H_1\ldots CX_{n+1\to n} H_n H_{n+1}\left|GHZ^{(n)}\right\rangle \otimes \left|0\right\rangle_{n+1}
\end{aligned} \qquad (6)$$

where the last line uses the fact that $H_n H_{n+1} CX_{n\to n+1} H_{n+1} H_n = CX_{n+1\to n}$. Then by

rearranging commutative gates we have:

$$\left|\psi^{(n+1)}\right\rangle = CX_{n+1 \to n} H_{n+1}\left(H_1...H_n \left|GHZ^{(n)}\right\rangle\right) \otimes \left|0\right\rangle_{n+1}$$

$$= CX_{n+1 \to n} H_{n+1}\left|\psi^{(n)}\right\rangle \otimes \left|0\right\rangle_{n+1} \tag{7}$$

which leads to Property 1 that $\left|\psi^{(n+1)}\right\rangle = \dfrac{1}{\sqrt{2}}\left(\left|\psi^{(n)}\right\rangle\left|0\right\rangle_{n+1} + \left|\tilde{\psi}^{(n)}\right\rangle\left|1\right\rangle_{n+1}\right)$.

**Property 2**: $\left|\psi^{(n)}\right\rangle$ and $\left|\tilde{\psi}^{(n)}\right\rangle$ together contain all computational basis states of the $n$-qubit space, i.e. $\left|00...0\right\rangle_{12...n}$ through $\left|11...1\right\rangle_{12...n}$, with equal coefficients.

Proof: Firstly we have $\left|\psi^{(3)}\right\rangle = \dfrac{1}{2}\left[\left(\left|00\right\rangle_{12} + \left|11\right\rangle_{12}\right)\left|0\right\rangle_3 + \left(\left|01\right\rangle_{12} + \left|10\right\rangle_{12}\right)\left|1\right\rangle_3\right]$ and $\left|\tilde{\psi}^{(3)}\right\rangle = \dfrac{1}{2}\left[\left(\left|00\right\rangle_{12} + \left|11\right\rangle_{12}\right)\left|1\right\rangle_3 + \left(\left|01\right\rangle_{12} + \left|10\right\rangle_{12}\right)\left|0\right\rangle_3\right]$, such that they together contain all $\left|000\right\rangle$ through $\left|111\right\rangle$ with the same coefficient $\dfrac{1}{2}$. Secondly, by Property 1, $\left|\psi^{(k+1)}\right\rangle = \dfrac{1}{\sqrt{2}}\left(\left|\psi^{(k)}\right\rangle\left|0\right\rangle_{k+1} + \left|\tilde{\psi}^{(k)}\right\rangle\left|1\right\rangle_{k+1}\right)$ and $\left|\tilde{\psi}^{(k+1)}\right\rangle = \dfrac{1}{\sqrt{2}}\left(\left|\psi^{(k)}\right\rangle\left|1\right\rangle_{k+1} + \left|\tilde{\psi}^{(k)}\right\rangle\left|0\right\rangle_{k+1}\right)$, so if we assume that $\left|\psi^{(k)}\right\rangle$ and $\left|\tilde{\psi}^{(k)}\right\rangle$ together contain all $\left|00...0\right\rangle_{12...k}$ through $\left|11...1\right\rangle_{12...k}$ of the $k$-qubit space with equal coefficients, then $\left|\psi^{(k+1)}\right\rangle$ and $\left|\tilde{\psi}^{(k+1)}\right\rangle$ together also contain all $\left|00...0\right\rangle_{12...k+1}$ through $\left|11...1\right\rangle_{12...k+1}$ of the $(k+1)$-qubit space with equal coefficients. Then by mathematical induction we have Property 2 proved.

**Proof for the Theorem**: In Step 3, $\left|\psi_D^{(n)}\right\rangle$ is created from $\left|\psi^{(n)}\right\rangle$ by applying $CX_{i \to iD}$ for each $q_i$ in $\left|\psi^{(n)}\right\rangle$ and $q_{iD} = \left|0\right\rangle$, where $i$ goes from 1 to $n-1$:

$$\left|\psi_D^{(n)}\right\rangle = CX_{1 \to 1D}...CX_{(n-1) \to (n-1)D}\left|\psi^{(n)}\right\rangle\left|0...0\right\rangle_{1D...(n-1)D} \tag{8}$$

where $\left|\psi_D^{(n)}\right\rangle$ simply has a duplicate $q_{iD}$ for each $q_i$ except $q_n$.

Here we define:

$$\left|\Phi_D^{(n)}\right\rangle = CX_{n \to nD}\left|\psi_D^{(n)}\right\rangle\left|0\right\rangle_{nD} = CX_{1 \to 1D}...CX_{n \to nD}\left|\psi^{(n)}\right\rangle\left|0...0\right\rangle_{1D...nD} \tag{9}$$

such that in $\left|\Phi_D^{(n)}\right\rangle$, $q_n$ also has a duplicate $q_{nD}$.

By Property 1, we transform $\left|\psi_D^{(n)}\right\rangle$ into:

$$\left|\psi_D^{(n)}\right\rangle = CX_{1\to 1D}...CX_{(n-1)\to(n-1)D}\left|\psi^{(n)}\right\rangle\left|0...0\right\rangle_{1D...(n-1)D}$$

$$= \frac{1}{\sqrt{2}}\left[\begin{array}{l}\left(CX_{1\to 1D}...CX_{(n-1)\to(n-1)D}\left|\psi^{(n-1)}\right\rangle\left|0...0\right\rangle_{1D...(n-1)D}\right)\left|0\right\rangle_n \\ + \left(CX_{1\to 1D}...CX_{(n-1)\to(n-1)D}\left|\tilde{\psi}^{(n-1)}\right\rangle\left|0...0\right\rangle_{1D...(n-1)D}\right)\left|1\right\rangle_n\end{array}\right] \tag{10}$$

$$= \frac{1}{\sqrt{2}}\left[\left|\Phi_D^{(n-1)}\right\rangle\left|0\right\rangle_n + \left|\tilde{\Phi}_D^{(n-1)}\right\rangle\left|1\right\rangle_n\right]$$

where $\left|\tilde{\Phi}_D^{(n-1)}\right\rangle = X_{n-1}X_{(n-1)D}\left|\Phi_D^{(n-1)}\right\rangle$ with $X_{n-1}$ and $X_{(n-1)D}$ negating the values of $q_{n-1}$ and $q_{(n-1)D}$. Now by Property 2, $\left|\Phi_D^{(n-1)}\right\rangle$ and $\left|\tilde{\Phi}_D^{(n-1)}\right\rangle$ together contain all computational basis states of the $(n-1)$-qubit space with duplicates: i.e. they together contain all $\left|00...0\right\rangle_{12...n-1}\otimes\left|00...0\right\rangle_{1D2D...(n-1)D}$ through $\left|11...1\right\rangle_{12...n-1}\otimes\left|11...1\right\rangle_{1D2D...(n-1)D}$, and the associated coefficients are all equal. To see a concrete example, for $n=3$ we have:

$$\left|\psi_D^{(3)}\right\rangle = \frac{1}{\sqrt{2}}\left[\left|\Phi_D^{(2)}\right\rangle\left|0\right\rangle_3 + \left|\tilde{\Phi}_D^{(2)}\right\rangle\left|1\right\rangle_3\right]$$

$$= \frac{1}{2}\left[\left(\left|00\right\rangle_{12}\left|00\right\rangle_{1D2D} + \left|11\right\rangle_{12}\left|11\right\rangle_{1D2D}\right)\left|0\right\rangle_3 + \left(\left|01\right\rangle_{12}\left|01\right\rangle_{1D2D} + \left|10\right\rangle_{12}\left|10\right\rangle_{1D2D}\right)\left|1\right\rangle_3\right] \tag{11}$$

where $\left|\Phi_D^{(2)}\right\rangle$ and $\left|\tilde{\Phi}_D^{(2)}\right\rangle$ together contain all computational basis states of the 2-qubit space with duplicates: $\left|00\right\rangle_{12}\otimes\left|00\right\rangle_{1D2D}$, $\left|01\right\rangle_{12}\otimes\left|01\right\rangle_{1D2D}$, $\left|10\right\rangle_{12}\otimes\left|10\right\rangle_{1D2D}$, $\left|11\right\rangle_{12}\otimes\left|11\right\rangle_{1D2D}$. Now we trace away $q_{1D}$, $q_{2D}$, and $q_3$ from $\left|\psi_D^{(3)}\right\rangle$, the reduced density state of $q_1$ and $q_2$ is:

$$\rho_r^{(2)} = \rho_{12} = Tr_{1D2D,3}\left(\left|\psi_D^{(3)}\right\rangle\left\langle\psi_D^{(3)}\right|\right) = \frac{1}{4}\left(\left|00\right\rangle\left\langle00\right| + \left|01\right\rangle\left\langle01\right| + \left|10\right\rangle\left\langle10\right| + \left|11\right\rangle\left\langle11\right|\right) = \frac{1}{4}I_4 \tag{12}$$

Here we emphasize that Eq. (12) only works because the duplicate qubits $q_{1D}$ and $q_{2D}$ in Eq. (11) have been added in Step 3 of the PSQE – if we remove them from Eq. (11) then the reduced density state would not be proportional to the identity.

This result can be generalized: when we consider Eq. (10) and trace away $q_{1D}$ through $q_{(n-1)D}$, and $q_n$, then the reduced density state of $q_1$ through $q_{n-1}$ is:

$$\rho_r^{(n-1)} = \rho_{1...n-1} = Tr_{1D...(n-1)D,n}\left(\left|\psi_D^{(n)}\right\rangle\left\langle\psi_D^{(n)}\right|\right) = \frac{1}{2^{n-1}}I_{2^{n-1}} \tag{13}$$

Once again Eq. (13) only works because the duplicate qubits $q_{1D}$ through $q_{(n-1)D}$ have been added in Step 3. The theorem has been proved. QED.

**Section S3. Proof for Theorem 2 in Section 3.1 of the main text.**

**Theorem 2**: If Eve is allowed to perform arbitrary unitary transformation before measuring the cipher qubits, then her "success probability" $P_s$ of inferring $q_n$'s value correctly when averaged over all possible key configurations satisfy:

$$\frac{1}{2} - \left(\frac{\sqrt{2}}{2}\right)^{n+1} = P_{\min} \leq P_s \leq P_{\max} = \frac{1}{2} + \left(\frac{\sqrt{2}}{2}\right)^{n+1} \tag{14}$$

**Proof**:

By Theorem 1, the reduced density state of the cipher qubits is proportional to the identity and invariant upon arbitrary unitary transformation, so Eve will not gain any information by observing the measurement statistics of the cipher qubits. What she can do is to use the values measured from the cipher qubits to infer $q_n$'s value. For example, Eve can guess $q_n = 0$ if $\bigoplus_{i=1}^{n-1} q_i = 0$ is measured, and define $P_s$ as the conditional probability:

$$P_s = P\left(q_n = 0 \left| \bigoplus_{i=1}^{n-1} q_i = 0\right.\right) \tag{15}$$

Now by Bayes' rule:

$$P\left(q_n = 0 \left| \bigoplus_{i=1}^{n-1} q_i = 0\right.\right) \cdot P\left(\bigoplus_{i=1}^{n-1} q_i = 0\right) = P\left(\bigoplus_{i=1}^{n-1} q_i = 0 \left| q_n = 0\right.\right) \cdot P\left(q_n = 0\right) \tag{16}$$

By Theorem 1 and the form of $\left|\psi_D^{(n)}\right\rangle$, $P\left(\bigoplus_{i=1}^{n-1} q_i = 0\right) = P\left(q_n = 0\right) = \frac{1}{2}$ always, so we have:

$$P_s = P\left(q_n = 0 \left| \bigoplus_{i=1}^{n-1} q_i = 0\right.\right) = P\left(\bigoplus_{i=1}^{n-1} q_i = 0 \left| q_n = 0\right.\right) \tag{17}$$

Now what if Eve measures $\bigoplus_{i=1}^{n-1} q_i = 1$ ? She can just guess $q_n = 1$, and define $P_s = P\left(q_n = 1 \left| \bigoplus_{i=1}^{n-1} q_i = 1\right.\right)$. To see that this alternative definition is equivalent to Eq. (15), we use the Bayes' rule in a similar manner to Eq. (16) and find:

$$P\left(q_n = 0 \middle| \bigoplus_{i=1}^{n-1} q_i = 0\right) = P\left(\bigoplus_{i=1}^{n-1} q_i = 0 \middle| q_n = 0\right)$$

$$= 1 - P\left(\bigoplus_{i=1}^{n-1} q_i = 1 \middle| q_n = 0\right)$$

$$= 1 - P\left(q_n = 0 \middle| \bigoplus_{i=1}^{n-1} q_i = 1\right) \quad (18)$$

$$= 1 - \left[1 - P\left(q_n = 1 \middle| \bigoplus_{i=1}^{n-1} q_i = 1\right)\right]$$

$$= P\left(q_n = 1 \middle| \bigoplus_{i=1}^{n-1} q_i = 1\right)$$

So indeed, $P\left(q_n = 1 \middle| \bigoplus_{i=1}^{n-1} q_i = 1\right) = P\left(q_n = 0 \middle| \bigoplus_{i=1}^{n-1} q_i = 0\right)$, and thus we can, without loss of generality, use Eq. (17) to evaluate $P_s$ for Eve. In particular, we consider $P\left(\bigoplus_{i=1}^{n-1} q_i = 0 \middle| q_n = 0\right)$, the conditional probability of the cipher qubits summing to 0, given that $q_n$ is already measured to be 0.

Because $q_n$ is already 0, we project $\left|\psi_D^{(n)}\right\rangle$ in Eq. (10) into the half state associated with $\left|0\right\rangle_n$ to get $\left|\Phi_D^{(n-1)}\right\rangle$, and then trace away the duplicate qubits $q_{1D}$ through $q_{(n-1)D}$ to get the reduced density state $\rho^{(n-1)}$ containing an equal mixture of all computational basis states of the ($n$-1)-qubit space that satisfy $\bigoplus_{i=1}^{n-1} q_i = 0$. Suppose Alice applies no rotation to the cipher qubits and Eve knows it, we always measure $\bigoplus_{i=1}^{n-1} q_i = 0$ from $\rho^{(n-1)}$ and $P_s = P\left(\bigoplus_{i=1}^{n-1} q_i = 0 \middle| q_n = 0\right)$ would be 1: this means Eve would always guess $q_n$ correctly. However, when Alice does apply Hadamard gates to the cipher qubits according to the key **k**, without knowing **k**, Eve has to average over all possible key configurations for the cipher qubits. This means that for Eve, the effective density state $\rho_e^{(n-1)}$ of the cipher qubits has to include all the basis states satisfying $\bigoplus_{i=1}^{n-1} q_i = 0$ and their associated key-variants due to possible Hadamard rotations. For example, for $n-1=2$, $\rho_e^{(2)}$ is an equal mixture of $\rho(00)$ composed of $|00\rangle$ and its three key-variants $|0+\rangle$, $|+0\rangle$, $|++\rangle$, plus $\rho(11)$ composed of $|11\rangle$ and its three key-variants $|1-\rangle$, $|-1\rangle$, $|--\rangle$:

$$\rho_e^{(2)} = \frac{1}{2}\Big[ \rho(00) + \rho(11) \Big]$$

$$\left( \begin{array}{l} \rho(00) = \dfrac{1}{4}\big(|00\rangle\langle 00| + |0+\rangle\langle 0+| + |+0\rangle\langle +0| + |++\rangle\langle ++|\big) \\[2mm] \rho(11) = \dfrac{1}{4}\big(|11\rangle\langle 11| + |1-\rangle\langle 1-| + |-1\rangle\langle -1| + |--\rangle\langle --|\big) \end{array} \right) \tag{19}$$

Similarly, for $n-1 = 3$ we have:

$$\rho_e^{(3)} = \frac{1}{4}\Big[ \rho(000) + \rho(011) + \rho(110) + \rho(101) \Big]$$

$$\left( \begin{array}{l} \rho(000) = \dfrac{1}{8}\left( \begin{array}{l} |000\rangle\langle 000| + |00+\rangle\langle 00+| + |0+0\rangle\langle 0+0| + |0++\rangle\langle 0++| \\ +|+00\rangle\langle +00| + |+0+\rangle\langle +0+| + |++0\rangle\langle ++0| + |+++\rangle\langle +++| \end{array} \right) \\[4mm] \rho(011) = \dfrac{1}{8}\left( \begin{array}{l} |011\rangle\langle 011| + |01-\rangle\langle 01-| + |0-1\rangle\langle 0-1| + |0--\rangle\langle 0--| \\ +|+11\rangle\langle +11| + |+1-\rangle\langle +1-| + |+-1\rangle\langle +-1| + |+--\rangle\langle +--| \end{array} \right) \\[4mm] \rho(110) = \dfrac{1}{8}\left( \begin{array}{l} |110\rangle\langle 110| + |11+\rangle\langle 11+| + |1-0\rangle\langle 1-0| + |1-+\rangle\langle 1-+| \\ +|-10\rangle\langle -10| + |-1+\rangle\langle -1+| + |--0\rangle\langle --0| + |--+\rangle\langle --+| \end{array} \right) \\[4mm] \rho(101) = \dfrac{1}{8}\left( \begin{array}{l} |101\rangle\langle 101| + |10-\rangle\langle 10-| + |1+1\rangle\langle 1+1| + |1+-\rangle\langle 1+-| \\ +|-01\rangle\langle -01| + |-0-\rangle\langle -0-| + |-+1\rangle\langle -+1| + |-+-\rangle\langle -+-| \end{array} \right) \end{array} \right) \tag{20}$$

where $\rho(q_1 q_2 q_3)$ includes all $2^3 = 8$ possible key-variants of $|q_1 q_2 q_3\rangle$.

For the general case:

$$\rho_e^{(n-1)} = \frac{1}{2^{n-2}} \sum_{q_1 \oplus \ldots \oplus q_{n-1} = 0} \rho(q_1 \ldots q_{n-1}) \tag{21}$$

where $\rho(q_1 \ldots q_{n-1})$ is defined as the equal mixture of all $2^{n-1}$ possible key-variants of $|q_1 \ldots q_{n-1}\rangle$, just like in Eqs. (19) and (20).

Now after an arbitrary unitary transformation $U$, $\rho_e^{(n-1)}$ becomes $U\rho_e^{(n-1)}U^\dagger$, and $P_s = P\left( \bigoplus_{i=1}^{n-1} q_i = 0 \,\middle|\, q_n = 0 \right)$ is just the probability of measuring the basis states satisfying $\bigoplus_{i=1}^{n-1} q_i = 0$ from $U\rho_e^{(n-1)}U^\dagger$:

$$P_s = P\left(\bigoplus_{i=1}^{n-1} q_i = 0 \,\middle|\, q_n = 0\right) = \sum_{q_1 \oplus \ldots \oplus q_{n-1} = 0} \left\langle q_1 \ldots q_{n-1} \,\middle|\, U \rho_e^{(n-1)} U^\dagger \,\middle|\, q_1 \ldots q_{n-1} \right\rangle \tag{22}$$

Note that $\rho_e^{(n-1)}$ is a Hermitian matrix, therefore Eq. (22) is essentially a sum of expectation values of the observable $\rho_e^{(n-1)}$ evaluated on some basis states rotated by the unitary transformation $U$. Given this understanding, we present Lemma 1, an extension of the variational principle:

**Lemma 1**: Let $H$ be an $N \times N$ Hermitian matrix, $\lambda_1 \le \lambda_2 \le \ldots \le \lambda_N$ be its eigenvalues, $|\phi_1\rangle$ through $|\phi_k\rangle$ be any $k$-subset ($1 \le k \le N$) of a complete set of orthonormal basis states, then we have: $\sum_{i=1}^{k} \lambda_i \le \sum_{i=1}^{k} \langle \phi_i | H | \phi_i \rangle \le \sum_{i=N-k+1}^{N} \lambda_i$.

**Proof**: (If not interested in the proof of Lemma 1, please go directly to Eq. (26).) Let $|\psi_1\rangle$ through $|\psi_N\rangle$ be eigenstates of $H$ with associated eigenvalues $\lambda_1$ through $\lambda_N$, then we can write each $|\phi_i\rangle$ as $|\phi_i\rangle = \sum_{j=1}^{N} c_{ij} |\psi_j\rangle$ and thus:

$$\sum_{i=1}^{k} \langle \phi_i | H | \phi_i \rangle = \sum_{i=1}^{k} \sum_{j=1}^{N} |c_{ij}|^2 \cdot \lambda_j = \sum_{j=1}^{N} \sum_{i=1}^{k} |c_{ij}|^2 \cdot \lambda_j = \sum_{j=1}^{N} \left(1 - \sum_{i=k+1}^{N} |c_{i,j}|^2\right) \cdot \lambda_j \tag{23}$$

where we have used the equality $\sum_{i=1}^{N} |c_{ij}|^2 = 1$. Next because $\lambda_1 \le \lambda_2 \le \ldots \le \lambda_N$, we have:

$$\begin{aligned}
\sum_{j=1}^{N} \left(1 - \sum_{i=k+1}^{N} |c_{i,j}|^2\right) \cdot \lambda_j &\ge \sum_{j=1}^{k-1} \left(1 - \sum_{i=k+1}^{N} |c_{i,j}|^2\right) \cdot \lambda_j + \sum_{j=k}^{N} \left(1 - \sum_{i=k+1}^{N} |c_{i,j}|^2\right) \cdot \lambda_k \\
&= \sum_{j=1}^{k-1} \left(1 - \sum_{i=k+1}^{N} |c_{i,j}|^2\right) \cdot \lambda_j + \left[(N-k+1) - \sum_{j=k}^{N} \sum_{i=k+1}^{N} |c_{i,j}|^2\right] \cdot \lambda_k \\
&= \sum_{j=1}^{k-1} \left(1 - \sum_{i=k+1}^{N} |c_{i,j}|^2\right) \cdot \lambda_j + \left[(N-k+1) - \sum_{i=k+1}^{N} \sum_{j=k}^{N} |c_{i,j}|^2\right] \cdot \lambda_k \\
&= \sum_{j=1}^{k-1} \left(1 - \sum_{i=k+1}^{N} |c_{i,j}|^2\right) \cdot \lambda_j + \left[(N-k+1) - \sum_{i=k+1}^{N} \left(1 - \sum_{j=1}^{k-1} |c_{i,j}|^2\right)\right] \cdot \lambda_k
\end{aligned} \tag{24}$$

where in the last line we have used the equality $\sum_{j=1}^{N} |c_{ij}|^2 = 1$. Then rearranging some terms we have:

$$\sum_{j=1}^{N}\left(1-\sum_{i=k+1}^{N}\left|c_{i,j}\right|^{2}\right)\cdot\lambda_{j} \geq \sum_{j=1}^{k-1}\left(1-\sum_{i=k+1}^{N}\left|c_{i,j}\right|^{2}\right)\cdot\lambda_{j}+\left[\left(N-k+1\right)-\sum_{i=k+1}^{N}\left(1-\sum_{j=1}^{k-1}\left|c_{i,j}\right|^{2}\right)\right]\cdot\lambda_{k}$$

$$=\sum_{j=1}^{k-1}\lambda_{j}-\sum_{j=1}^{k-1}\sum_{i=k+1}^{N}\left|c_{i,j}\right|^{2}\cdot\lambda_{j}+\left[\left(N-k+1\right)-\left(N-k\right)+\sum_{i=k+1}^{N}\sum_{j=1}^{k-1}\left|c_{i,j}\right|^{2}\right]\cdot\lambda_{k}$$

$$=\sum_{j=1}^{k}\lambda_{j}+\sum_{i=k+1}^{N}\sum_{j=1}^{k-1}\left|c_{i,j}\right|^{2}\cdot\left(\lambda_{k}-\lambda_{j}\right)$$

$$\geq \sum_{j=1}^{k}\lambda_{j}$$

$$(25)$$

So indeed $\sum_{i=1}^{k}\left\langle\phi_{i}\left|H\right|\phi_{i}\right\rangle\geq\sum_{j=1}^{k}\lambda_{j}$. By symmetry, the inequality of $\sum_{i=1}^{k}\left\langle\phi_{i}\left|H\right|\phi_{i}\right\rangle\leq\sum_{i=N-k+1}^{N}\lambda_{i}$ can be proved by exactly the same procedure. So Lemma 1 has been proved.

By Lemma 1 and Eq. (22), given $\lambda_{1}\leq\lambda_{2}\leq...\leq\lambda_{2^{n-1}}$ are the eigenvalues of $\rho_{e}^{(n-1)}$, we have:

$$\sum_{i=1}^{2^{n-2}}\lambda_{i}\ \leq\ P_{s}\ \leq\ \sum_{i=2^{n-2}+1}^{2^{n-1}}\lambda_{i} \qquad (26)$$

Next to evaluate the eigenvalues of $\rho_{e}^{(n-1)}$, we first consider the simple case of Eq. (19). By defining $\rho\left(0\right)=\frac{1}{2}\left(\left|0\right\rangle\left\langle0\right|+\left|+\right\rangle\left\langle+\right|\right)$ and $\rho\left(1\right)=\frac{1}{2}\left(\left|1\right\rangle\left\langle1\right|+\left|-\right\rangle\left\langle-\right|\right)$, we find:

$$\rho\left(00\right)=\frac{1}{4}\left(\left|00\right\rangle\left\langle00\right|+\left|0+\right\rangle\left\langle0+\right|+\left|+0\right\rangle\left\langle+0\right|+\left|++\right\rangle\left\langle++\right|\right)$$

$$=\frac{1}{2}\left(\left|0\right\rangle\left\langle0\right|+\left|+\right\rangle\left\langle+\right|\right)\otimes\frac{1}{2}\left(\left|0\right\rangle\left\langle0\right|+\left|+\right\rangle\left\langle+\right|\right)$$

$$=\rho\left(0\right)\otimes\rho\left(0\right)$$

$$\rho\left(11\right)=\frac{1}{4}\left(\left|11\right\rangle\left\langle11\right|+\left|1-\right\rangle\left\langle1-\right|+\left|-1\right\rangle\left\langle-1\right|+\left|--\right\rangle\left\langle--\right|\right)$$

$$=\frac{1}{2}\left(\left|1\right\rangle\left\langle1\right|+\left|-\right\rangle\left\langle-\right|\right)\otimes\frac{1}{2}\left(\left|1\right\rangle\left\langle1\right|+\left|-\right\rangle\left\langle-\right|\right)$$

$$=\rho\left(1\right)\otimes\rho\left(1\right)$$

$$(27)$$

Similarly, if we consider the case in Eq. (20), for $\rho\left(000\right)$ we have:

$$\rho(000) = \frac{1}{8}\begin{pmatrix} |000\rangle\langle000| + |00+\rangle\langle00+| + |0+0\rangle\langle0+0| + |0++\rangle\langle0++| \\ + |+00\rangle\langle+00| + |+0+\rangle\langle+0+| + |++0\rangle\langle++0| + |+++\rangle\langle+++| \end{pmatrix}$$

$$= \frac{1}{2}\left(|0\rangle\langle0| + |+\rangle\langle+|\right) \otimes \frac{1}{2}\left(|0\rangle\langle0| + |+\rangle\langle+|\right) \otimes \frac{1}{2}\left(|0\rangle\langle0| + |+\rangle\langle+|\right) \qquad (28)$$

$$= \rho(0) \otimes \rho(0) \otimes \rho(0)$$

By inspecting Eqs. (27) and (28), we see the structures of $\rho(00)$ and $\rho(000)$ are essentially a binomial expansion of $(a+b)^{\otimes n-1}$, with $a = \frac{1}{2}|0\rangle\langle0|$, $b = \frac{1}{2}|+\rangle\langle+|$, and $a+b = \rho(0)$. Similarly, $\rho(11)$ is just a binomial expansion of $(c+d)^{\otimes n-1}$, with $c = \frac{1}{2}|1\rangle\langle1|$, $d = \frac{1}{2}|-\rangle\langle-|$, and $c+d = \rho(1)$. Now similarly by inspecting the structures in Eq. (20), we have:

$$\rho(011) = \rho(0) \otimes \rho(1) \otimes \rho(1)$$
$$\rho(110) = \rho(1) \otimes \rho(1) \otimes \rho(0) \qquad (29)$$
$$\rho(101) = \rho(1) \otimes \rho(0) \otimes \rho(1)$$

In Eq. (29), $\rho(q_1 q_2 q_3) = \rho(q_1) \otimes \rho(q_2) \otimes \rho(q_3)$, and this result also extends to the general case because the structure of $\rho(q_1...q_{n-1})$ is essentially the same, and we have:

$$\rho(q_1...q_{n-1}) = \rho(q_1) \otimes ... \otimes \rho(q_{n-1}) \qquad (30)$$

Consequently, Eq. (21) becomes:

$$\rho_e^{(n-1)} = \frac{1}{2^{n-2}} \sum_{q_1 \oplus ... \oplus q_{n-1} = 0} \rho(q_1) \otimes ... \otimes \rho(q_{n-1}) \qquad (31)$$

Now the summation in Eq. (31) goes over all outcomes satisfying $\bigoplus_{i=1}^{n-1} q_i = 0$, so there can only be an even number of $\rho(1)$'s in the summand, and thus the summation can be further transformed, by the binomial expansion again we have:

$$\rho_e^{(n-1)} = \frac{1}{2^{n-2}} \sum_{q_1 \oplus ... \oplus q_{n-1} = 0} \rho(q_1) \otimes ... \otimes \rho(q_{n-1})$$

$$= \frac{1}{2^{n-3}}\left(\left[\rho(0) + \rho(1)\right]^{\otimes n-1} + \left[\rho(0) - \rho(1)\right]^{\otimes n-1}\right) \qquad (32)$$

where two binomial expansions are added such that the terms involving an odd number of $\rho(1)$'s are cancelled, while the terms involving an even number of $\rho(1)$'s remain.

Next we recognize that $\left[\rho(0),\rho(1)\right]=0$, thus they can be simultaneously diagonalized. After some simple calculations we have:

$$\rho(0)|\phi_1\rangle=\mu_1|\phi_1\rangle,\quad \rho(0)|\phi_2\rangle=\mu_2|\phi_2\rangle,\quad \rho(1)|\phi_1\rangle=\mu_2|\phi_1\rangle,\quad \rho(1)|\phi_2\rangle=\mu_1|\phi_2\rangle$$

$$|\phi_1\rangle=\frac{\sqrt{2+\sqrt{2}}}{2}|0\rangle+\frac{\sqrt{2-\sqrt{2}}}{2}|1\rangle,\quad |\phi_2\rangle=\frac{\sqrt{2-\sqrt{2}}}{2}|0\rangle-\frac{\sqrt{2+\sqrt{2}}}{2}|1\rangle \tag{33}$$

$$\mu_1=\frac{2+\sqrt{2}}{4},\quad \mu_2=\frac{2-\sqrt{2}}{4}$$

By Eqs. (32) and (33) we can use the tensor products of $|\phi_1\rangle$ and $|\phi_2\rangle$ to create an entire eigenbasis of $\rho_e^{(n-1)}$: e.g. for $n-1=3$ we can have eigenstates like $|\psi_1\rangle=|\phi_1\rangle\otimes|\phi_1\rangle\otimes|\phi_1\rangle$, $|\psi_2\rangle=|\phi_1\rangle\otimes|\phi_1\rangle\otimes|\phi_2\rangle$, ... , $|\psi_8\rangle=|\phi_2\rangle\otimes|\phi_2\rangle\otimes|\phi_2\rangle$. In addition, there can only be two unique eigenvalues: when an eigenstate $|\psi\rangle$ contains $k$ $|\phi_2\rangle$'s in the tensor product, we have:

$$\rho_e^{(n-1)}|\psi\rangle=\frac{1}{2^{n-3}}\left[\left(\mu_1+\mu_2\right)^{n-1-k}\left(\mu_2+\mu_1\right)^k+\left(\mu_1-\mu_2\right)^{n-1-k}\left(\mu_2-\mu_1\right)^k\right]|\psi\rangle \tag{34}$$

So the eigenvalue is:

$$\lambda=\frac{1}{2^{n-3}}\left[\left(\mu_1+\mu_2\right)^{n-1}+\left(-1\right)^k\left(\mu_1-\mu_2\right)^{n-1}\right]$$

$$=\frac{1}{2^{n-3}}\left[1+\left(-1\right)^k\left(\frac{\sqrt{2}}{2}\right)^{n-1}\right] \tag{35}$$

So the lowest half eigenvalues are all $\dfrac{1}{2^{n-3}}\left[1-\left(\dfrac{\sqrt{2}}{2}\right)^{n-1}\right]$ and highest half eigenvalues are all $\dfrac{1}{2^{n-3}}\left[1+\left(\dfrac{\sqrt{2}}{2}\right)^{n-1}\right]$, therefore by Eq. (26) we have:

$$\frac{1}{2}-\left(\frac{\sqrt{2}}{2}\right)^{n+1}=\sum_{i=1}^{2^{n-2}}\lambda_i \leq P_s \leq \sum_{i=2^{n-2}+1}^{2^{n-1}}\lambda_i=\frac{1}{2}+\left(\frac{\sqrt{2}}{2}\right)^{n+1} \tag{36}$$

This is exactly Eq. (14), so Theorem 2 has been proved. QED.